# Neuronal Response Impedance Mechanism Implementing Cooperative Networks with Low Firing Rates and μs Precision

**Roni Vardi[1][†], Amir Goldental[2][†], Hagar Marmari[1], Haya Brama[1], Edward Stern[1,3], Shira Sardi[1,2], Pinhas Sabo[1,2] and Ido Kanter[1,2,*]**

[1]Gonda Interdisciplinary Brain Research Center and the Goodman Faculty of Life Sciences, Bar-Ilan University, Ramat-Gan 52900, Israel

[2]Department of Physics, Bar-Ilan University, Ramat-Gan 52900, Israel

[3]MassGeneral Institute for Neurodegenerative Disease, Department of Neurology, Massachusetts General Hospital, MA, USA

†These authors contributed equally to this work.

*Correspondence: ido.kanter@biu.ac.il

**Abstract**

**Realizations of low firing rates in neural networks usually require globally balanced distributions among excitatory and inhibitory links, while feasibility of temporal coding is limited by neuronal millisecond precision. We show that cooperation, governing global network features, emerges through nodal properties, as opposed to link distributions. Using *in vitro* and *in vivo* experiments we demonstrate microsecond precision of neuronal response timings under low stimulation frequencies, whereas moderate frequencies result in a chaotic neuronal phase characterized by degraded precision. Above a critical stimulation frequency, which varies among neurons, response failures were found to emerge stochastically such that the neuron functions as a low pass filter, saturating the average inter-spike-interval. This intrinsic neuronal response impedance mechanism leads to cooperation on a network level, such that firing rates are suppressed towards the lowest neuronal critical frequency simultaneously with neuronal microsecond precision. Our findings open up opportunities of controlling global features of network dynamics through few nodes with extreme properties.**

## 1. INTRODUCTION

The attempt to understand high cognitive functionalities and cooperative activities of neurons within a network results in many open questions. One question is which mechanism underlies the extremely low firing rates, few Hertz, of neurons comprising a network of threshold units, as a single neuron is capable of much higher firing frequencies(Amit and Brunel, 1997;Shafi et al., 2007;He et al., 2010;O'Connor et al., 2010). The second puzzle is the realization of precise neuronal response timings on a network level(VanRullen et al., 2005;Butts et al., 2007;Panzeri et al., 2010), whereas their variations are typically in the order of several milliseconds(Lass and Abeles, 1975;Mainen and Sejnowski, 1995;Schoppa and Westbrook, 1999;Foust et al., 2010). This realization is a prerequisite for the feasibility of temporal codes, which might play a role in brain functionalities. In this work we demonstrate that network low firing rates and neuronal precise response timings both stem from a single neuronal property, the neuronal response impedance mechanism, which counter-intuitively leads to cooperation among individual neurons comprising a network. The global features of network dynamics are governed by the distribution tail of the nodal properties, neuronal properties, rather than by specific distributions of the network links, the synapses.

The phenomenon of extremely low firing rates on the network level, about 1-3 Hz(Shafi et al., 2007;O'Connor et al., 2010), requires a balance between two opposing trends, spike birth and death. The rate of evoked spikes in an excitatory network is expected to constantly increase if each neuron excites several other neurons. This spike birth trend is moderated by a death trend, eliminating spikes by inhibitory synapses or weakening excitatory synapses to be sub-threshold(Turrigiano, 2008;Daqing et al., 2011). In case the birth and death trends are not precisely balanced, the network firing rate either diverges toward extremely high firing frequencies or practically vanishes(Turrigiano and Nelson, 2004;Chih et al., 2005). This balance can be theoretically achieved by several predefined synaptic designs depending on the topology of the network, such as a wide distribution of excitatory synaptic strengths balanced by a fraction of inhibitory synapses(van Vreeswijk and Sompolinsky, 1998;Brunel, 2000;Song et al., 2000;Vogels and Abbott, 2009;Vogels et al., 2011;Teramae et al., 2012;Spiegel et al., 2014). A second approach relies on modular networks(Diesmann et al., 1999;Litvak et al., 2003;Kumar et al., 2010;Rad et al., 2012). The firing activity of each module is maintained by intra-excitatory synapses, whereas low firing rates are achieved by synaptic inhibition among the modules, resulting in their alternate firing.

The second phenomenon, the realization of consistent and reliable neuronal response timings, which is a prerequisite for a possible realization and usefulness of temporal coding, is limited by the least precise building block of the network, i.e. neurons and synapses. Synaptic conductance is reproducible with accuracy of dozens of microseconds(Csicsvari et al., 1998;Doyle and Andresen, 2001;Rodríguez-Moreno et al., 2011), which is meaningful for temporal coding only if the temporal precision of a neuron is comparable. Indeed, there is some experimental verification of specific neurons functioning with very fine temporal resolution(Carr, 1993;Agmon-Snir et al., 1998;Kayser et al., 2010); nevertheless, the temporal resolution of cortical neurons is unclear. Neuronal precision can be experimentally determined by the neuronal response latency (NRL), which reflects the internal dynamics of the neuron and is measured as the time-lag between a stimulation and its corresponding evoked spike(Wagenaar et al., 2004;De Col et al., 2008;Gal et al., 2010;Vardi et al., 2012). This quantity dynamically varies by several milliseconds and is expected to accumulate to dozens of milliseconds in a neuronal chain and even more in recurrent networks(Vardi et al., 2013a;Vardi et al., 2013b). Hence, a necessary prerequisite for the realization of consistent and precise temporal coding on a network level is in question. Nevertheless, the possible relevance and usefulness of temporal coding to brain functionalities are beyond the scope of our work.

In this work we study experimentally, *in vitro* and *in vivo*, the neuronal response impedance, i.e. the statistics of the neuronal response timings and response failures to different stimulation frequencies. Specifically we find that under low stimulation rates neuronal responses are precise up to several µs, while higher stimulation rates lead to an increased imprecision and response failures. Simulations of large networks as well as theoretical arguments supported by experimental evidences indicate that those neuronal response failures lead to low firing rates, even in excitatory networks. Hence, the dynamics on a network level lead to the coexistence of low firing rates and µs precision. It calls for the reexamination of the role of inhibition in the brain activity and its contribution to the stabilization of the network low firing rates. This work also implies that it is highly possible that neurons operate under µs precision, which is a prerequisite for the formation of reliable temporal coding.

## 2. MATERIALS AND METHODS

### 2.1 Animals

All procedures were in accordance with the National Institutes of Health Guide for the Care and Use of Laboratory Animals and Bar-Ilan University Guidelines for the Use and Care of Laboratory Animals in Research and were approved and supervised by the Institutional Animal Care and Use Committee.

### 2.2 *In vitro* experiments

**2.2.1 Culture preparation.** Cortical neurons were obtained from newborn rats (Sprague-Dawley) within 48 h after birth using mechanical and enzymatic procedures(Vardi et al., 2012;2013c). The cortical tissue was digested enzymatically with 0.05% trypsin solution in phosphate-buffered saline (Dulbecco's PBS) free of calcium and magnesium, and supplemented with 20 mM glucose, at 37°C. Enzyme treatment was terminated using heat-inactivated horse serum, and cells were then mechanically dissociated. The neurons were plated directly onto substrate-integrated multi-electrode arrays (MEAs) and allowed to develop functionally and structurally mature networks over a time period of 2-4 weeks *in vitro*, prior to the experiments. The number of plated neurons in a typical network was in the order of 1,300,000, covering an area of about 380 mm$^2$ (i.e., ~0.32 neurons in 100 μm$^2$). The preparations were bathed in minimal essential medium (MEM-Earle, Earle's Salt Base without L-Glutamine) supplemented with heat-inactivated horse serum (5%), glutamine (0.5 mM), glucose (20 mM), and gentamicin (10 g/ml), and maintained in an atmosphere of 37°C, 5% $CO_2$ and 95% air in an incubator as well as during the electrophysiological measurements.

**2.2.2 Synaptic blockers.** All *in vitro* experiments, except for the experiments shown in **Figure 9**, were conducted on cultured cortical neurons that were functionally isolated from their network by a pharmacological block of glutamatergic and GABAergic synapses. For each culture 20 μl of a cocktail of synaptic blockers was used, consisting of 10 μM CNQX (6-cyano-7-nitroquinoxaline-2,3-dione), 80 μM APV (amino-5-phosphonovaleric acid) and 5 μM bicuculline. This cocktail did not block the spontaneous network activity completely, but rather made it sparse. At least one hour was allowed for stabilization of the effect.

**2.2.3 Stimulation and recording.** An array of 60 Ti/Au/TiN extracellular electrodes, 30 μm in diameter, and spaced either 200 or 500 μm from each other (Multi-Channel Systems, Reutlingen, Germany) were used. The insulation layer (silicon nitride) was pre-treated with polyethyleneimine (0.01% in 0.1 M Borate buffer solution). A commercial setup (MEA2100-2x60-headstage, MEA2100-interface board, MCS, Reutlingen, Germany) for recording and analyzing data from two 60-electrode MEAs was used, with integrated data acquisition from 120 MEA electrodes and 8 additional analog channels, integrated filter amplifier and 3-channel current or voltage stimulus generator (for each 60 electrode array). Mono-phasic square voltage pulses typically in the range of [-800, -500] mV and [60, 200] μs were applied through extracellular electrodes. Each channel was sampled at a frequency of 50k samples/s, thus the changes in the neuronal response latency were measured at a resolution of 20 μs.

**2.2.4 Cell selection.** Each node was represented by a stimulation source (source electrode) and a target for the stimulation – the recording electrode (target electrode). These electrodes (source and target) were selected as the ones that evoked well-isolated, well-formed spikes and reliable response with a high signal-to-noise ratio. This examination was done with a stimulus intensity of -800 mV with a duration of 200 μs using 30 repetitions at a rate of 5 Hz followed by 1200 repetitions at a rate of 10 Hz.

**2.2.5 Data analysis.** Analyses were performed in a Matlab environment (MathWorks, Natwick, MA, USA). The reported results were confirmed based on at least eight experiments each, using different sets of neurons and several tissue cultures.

Action potentials at experiments of the real-time adaptive algorithm for the stabilization of the neuronal response latency around a predefined latency were detected on-line by threshold crossing, using a detection window of typically 2-15 ms following the beginning of an electrical stimulation(Wagenaar et al., 2004).

In order to overcome the temporal precision of 20 μs determined by the maximal sampling rate of our recording device (50 kHz), in all *in vitro* experiments, except the above-mentioned adaptive algorithm, the following linear interpolation method for spike detection was used. For a given threshold crossing ($v_{threshold}$), we identify the two nearby sampling points: $(t_1, V_1)$ and $(t_2, V_2)$ where $V_1 \geq V_{threshold}$, $V_2 < V_{threshold}$ and $t_2 = t_1 + 20$ μs. Using linear interpolation between these two sampling points, the threshold crossing time, $t_{threshold}$, is estimated as (see also **Figure 6**)

$$t_{threshold} = t_1 + \frac{V_{threshold} - V_1}{V_2 - V_1} 20 \text{ μs}$$

The neuronal response latency was then calculated as the duration from the beginning of a stimulation to $t_{threshold}$.

### 2.3 Simulations

**2.3.1 Methods of simulation.** We simulated a network of 2000 excitatory leaky integrate and fire neurons (N=2000). The voltage $V'_i(t)$ of neuron i (i ∈ [1,N]), is given by the equation:

$$\frac{dV'_i}{dt} = -\frac{V'_i - V_{stable}}{\tau} + \sum_{j=1}^{2000} J'_{ji} \sum_{\substack{t' \in firing \\ times\ of \\ neuron\ j}} \delta\big(t - t' - D_{ji}\big) + J' \sum_{\substack{t' \in times \\ of\ external \\ stimulation \\ of\ neuron\ i}} \delta\big(t - t'\big)$$

with $V_{stable}$=-70 mV and a threshold of $V_{threshold}$=-54 mV, $\tau$=20 ms is the membrane time constant, $J'_{ji}$ is the connection strength from neuron j to neuron i (see connectivity section) and $D_{ji}$ is the time delay from an evoked spike of neuron j to the stimulation of neuron i and is randomly chosen from a flat distribution $U(6, 9.5)$ ms.

Under the variable substitution V'= ($V_{threshold}$-$V_{stable}$)·V+$V_{stable}$, $J'_{ji}$=16$J_{ji}$ mV and J'=16J mV the equation of the voltage, V, becomes now

$$\frac{dV_i}{dt} = -\frac{V_i}{\tau} + \sum_{j=1}^{2000} J_{ji} \sum_{\substack{t' \in firing \\ times\ of \\ neuron\ j}} \delta\big(t - t' - D_{ji}\big) + J \sum_{\substack{t' \in times \\ of\ external \\ stimulations \\ of\ neuron\ i}} \delta\big(t - t'\big).$$

For simplicity, we use this version of equation in the manuscript, since under this scaling $V_{threshold}$=1, $V_{stable}$=0, J>1 is above threshold and J<1 is bellow threshold. Nevertheless, results are the same for both equations.

The initial voltage is $V_i(t = 0) = 0.5\ \forall\ i$ and the integration is done using the Euler method with 0.05 ms time step.

If $V_i$ crosses the threshold, 1, the neuron may fire (see response failure section). If the neuron fires the voltage is reset to -0.5 after a refractory period of 2 ms, in which the neuron is inactive, does not respond to new stimulations. In the case of a response failure, the voltage is set to 0.2 without a refractory period.

**2.3.2 Connectivity.** Initially all connections are set to zero, i.e. all $J_{ji}$=0. First, we go over all neurons and randomly select a post-synaptic neuron for each one. Each neuron can be selected as a post-synaptic neuron only once, and a neuron cannot be connected to itself, i.e. $J_{ii}$=0. As a result of this procedure each neuron has only one pre-synaptic neuron and only one post-synaptic neuron. Next, we select with a probability of 0.1/N, unless stated differently, additional above-threshold connections. The strength of all above-threshold connections is set to $J_{ji}$=2, i.e. $J'_{ji}$=32 mV.

**2.3.3 Response failure.** We define t(i,n) as the time neuron i crossed the threshold for the $n^{th}$ time. The response failure probability for the $n^{th}$ threshold crossing of neuron i is:

$$P_{fail}(i,n) = \frac{\sum_{k<n}\left(1 - \frac{t(i,k+1)-t(i,k)}{\tau_C(i)}\right)e^{-\alpha(n-k+1)}}{\sum_{k<n}e^{-\alpha(n-k+1)}} \qquad (1)$$

where $\alpha$ is a measure of the neuronal forgetfulness and is equal to 1.4, unless stated differently. $1/\tau_C(i)$ is the critical frequency, $f_C(i)$, of neuron i, randomly chosen for each neuron, described in the figures. Note that negative response failure is taken as zero and in the limiting case of a periodic stimulation pattern, with the frequency $f_{stim}$, the expected response failure is obtained,

$$P_{fail}(i,n) = 1 - \left(\frac{f_C(i)}{f_{stim}}\right) \qquad (2)$$

independent of n. The simulations on a network level are found to be independent of the initial conditions, which in the presented simulations were taken as $P_{fail}(i,0)=0$.

**2.3.4 Initial stimulations.** To start the activity of the network the following external stimulations are chosen: First we randomly generate the external stimulation times for each neuron through a Poisson process with a rate of 50 Hz. Each external stimulation has a survival probability of exp(-T/200), where T stands for the time of the external stimulation, measured in milliseconds. Only stimulations with T<1000 ms are taken into account, as after 1000 ms the network has a self-sustaining activity, without spontaneous firing. Results are found to be insensitive to different initial conditions, e.g. only a single stimulation to a single neuron at time 0.

**2.3.5 Frequency histograms.** The firing frequency of a neuron was determined as the number of times the neuron fired between 4 and 59 seconds divided by 55 seconds. The bin size for all histograms is 0.5 Hz (**Figures 11B-E,12**).

**2.3.6 Additional response failure probability.** In the simulations shown at **Figure 11D** the response failure probability is C+(1-C)·$P_{fail}$, where $P_{fail}$ is given by equation (1) and C is a constant. For stimulation frequencies below $f_C$, the response failure probability is C as $P_{fail}$ is zero. In **Figure 11D**, C=0.07.

**2.3.7 Spontaneous activity.** In the simulations shown at **Figure 11E** the probability for a spontaneous stimulation per integration time step is $5·10^{-5}$, resulting on the average in 1 spontaneous stimulation per second for each one of the neurons.

**2.3.8 Additional sub-threshold connections.** In the simulations shown at **Figure 12C** there are additional sub-threshold connections, where every $J_{ji}=0$ was changed to $J_{ji}=J_{sub}$ with a probability $P_{sub}$ ($J_{ii}=0 \ \forall \ i$).

### 2.4 *In vivo* experiments

**2.4.1 Surgery.** The experiments were performed on Sprague-Dawley or Wistar rats weighing 100–200 g, initially anaesthetized with urethane (1.25 g per kg body weight, intraperitoneal), and given hourly supplemental injections of ketamine–xylazine (30 mg per kg and 7 mg per kg, respectively, intramuscular). The animals' temperatures were maintained at 37 °C, while placed in a stereotaxic instrument. All animals continued to breathe without artificial respiration, and were suspended by ear-bars and a clamp at the base of the tail to minimize movements of the brain caused by breathing. The cisterna magna was opened to relieve intracranial pressure. Small (1 mm) holes were drilled in the skull to allow insertion of electrocorticogram electrodes, in addition to a A 8X8 mm craniotomy for insertion of the recording and stimulating electrodes over the prefrontal cortex, where the dura was removed.

**2.4.2 Electrophysiology.** Two stimulating electrodes (tapered tungsten wire) were inserted about 0.6 mm deep into the cortex, to approximately cortical layer IV. Their tips had fixed distances of about 0.2 mm. The intra-cellular recording electrode was inserted within a radius of 1.5 mm into a depth of 0.2-1.2 mm, as a glass micropipette filled with 2–4% biocytin (Sigma) dissolved in 1 M potassium acetate. Electrode resistances ranged from 30 to 100 MΩ. After insertion of both types of electrodes, the exposed cortex was covered with a low-melting-point paraffin wax to reduce brain pulsations. Stimulus current amplitudes used were between 50 to 700 μA, and the duration of current flow was 100-200 μs. Recordings were made using an active bridge amplifier and then filtered and digitized at a rate of 10 kHz. The experiments were performed on neurons with membrane potentials lower than -60 mV and action potentials higher than 0 mV.

## 3. RESULTS

### 3.1 Stabilization of the neuronal response latency

When a neuron is repeatedly stimulated its response latency, NRL, stretches gradually. This effect was demonstrated for a 30 Hz stimulation frequency (**Figure 1A**) using cultured cortical neurons functionally separated from their network by synaptic blockers and stimulated such that most stimulations resulted in an evoked spike (MATERIALS AND METHODS). The accumulated stretching over few hundreds of repeated stimulations is typically several milliseconds, comparable with the initial NRL (**Figure 1A**). This stretching terminates at the intermittent phase, where the average NRL remains constant and is accompanied by both large fluctuations that can exceed a millisecond and a non-negligible fraction of neuronal response failures (**Figure 1A**). The NRL increase is a fully reversible phenomenon(Marmari et al., 2014), which considerably decays after few seconds without stimulations.

The stretching of the NRL seems to prevent consistent response timings, required for the realization of temporal codes. Even at the intermittent phase, where the average NRL is stable, the emergence of large fluctuations and neuronal response failures opposes the realization of temporal codes. Hence, we examine the feasibility of NRL stabilization using three different stimulation scenarios.

The proof of concept for NRL stabilization is examined first under an adaptive external feedback scenario(Steingrube et al., 2010) (**Figure 1B**). It relies on the fact that the average stretching of the NRL per stimulation increases with the decrease of the time-lag between stimulations, and vice versa(Vardi et al., 2012). Accordingly, a real-time adaptive algorithm for the stabilization of the NRL around a predefined latency, $L_{ST}$, was experimentally tested, indicating a supreme stabilization measured by the standard deviation $\sigma \sim 16$ μs (**Figure 1C**). In order to maintain stabilization around $L_{ST}$ the time-lag between stimulations has to be adjusted continuously, however, after ~1000 stimulations the average $\tau$ stabilizes around 200 ms.

This proof of concept for NRL stabilization, without response failures, calls for a more natural realization of this phenomenon without adaptive external control. Using a fixed time-lag of ~200 ms between stimulations, equal to the average time-lag obtained under the external feedback algorithm (**Figure 1C**), stabilization around the same $L_{ST} \sim 5$ ms is obtained, with a comparable standard deviation of $\sigma \sim 14$ μs (**Figure 1C**), however the transient to stabilization is longer (**Figures 2A,B**). Results suggest that the stabilized NRL, $L_{ST}$, is primarily a function of the average time-lag between stimulations. This was confirmed through a more realistic scenario where the same neuron was stimulated following random time-lags characterized by the same average of ~200 ms (**Figure 1C**), resulting in a similar $L_{ST}$ with slightly larger fluctuations, $\sigma$. In this case of stimulation with random time-lags, the $L_{ST}$ is slightly reduced as a result of asymmetric fluctuations in the neuronal response to sudden decrease or increase of the time-lags between consecutive stimulations, forming momentarily depression or facilitations in the neuronal response latency (**Figures 2C,D**).

We examine the effect of an abrupt transition to a much higher stimulation frequency following different periods of stabilization around $L_{ST}$. This scenario is exemplified by the stabilization of the neuronal response latency around $L_{ST} \sim 9.7$ ms using $\tau = 110$ ms, followed by an abrupt change to much shorter time-lags between stimulations, $\tau = 25$ ms (**Figure 3A**). A comparison between the segments of the neuronal response latency profiles following the change in the stimulation frequency shows that they are fairly identical (**Figure 3B**), indicating a lack of dependency on the history of simulations. In contrast, when the same neuron is stimulated solely with $\tau = 25$ ms its neuronal response latency profile above $L_{ST}$ is clearly different (**Figures 3A,B**). Before the latency stabilizes, it is solely a function of a global quantity, the averaged time-lag between all previous stimulations constitute the current neuronal response latency stretching (Vardi et al., 2014). This behavior indicates neuronal long-term memory (Vardi et al., 2014), in contrast to the behavior after stabilization.

The silencing of long-term memory during the stabilized period of the neuronal response latency does not disable the ultra-fast neuronal plasticity. Specifically, a substantial decrease in $\tau$ results in a momentary decrease in the neuronal response latency, indicating ultra-fast plasticity in the form of facilitation (**Figures 3A,C**). For example, when $\tau$ is shortened from 110 to 25 ms, neuronal facilitation with an amplitude of $\sim 0.2$ ms is evident, and the subsequent latency profiles are independent of the duration of stabilization at $L_{ST}$ (**Figure 3C**). A similar facilitation, with an amplitude of $\sim 0.2$ ms, also occurs when $\tau$ is momentarily shortened from 110 to 25 ms once every 100 stimulations (**Figure 3D**). The continuation of the response latency profile is recovered immediately following the next stimulation, either during the latency stretching phase or at the stabilization around $L_{ST}$, indicating insensitivity of global neuronal response latency profiles to momentary leaps in $\tau$. The co-existence of these two features, silenced long-term neuronal memory and neuronal ultra-fast plasticity (**Figure 3A**), suggests a possible realization of reliable signaling in temporal codes (Vardi et al., 2014), especially when the neuronal response latency is stable.

### 3.2 Universal properties of the stable neuronal response latency

### 3.2.1 In vitro experiments

The NRL profiles differ among neurons in their total increase, average increase per stimulation and detailed profile forms. In case the NRL can be stabilized at any latency, are there common neuronal trends, independent of the detailed NRL profile? Specifically, can one find universal features characterizing the fluctuations around $L_{ST}$, $\sigma(L_{ST})$, and the average time-lag between stimulations required for stabilization around $L_{ST}$, $\tau(L_{ST})$? Stabilization close to the initial NRL requires $\tau$ in the order of seconds which rapidly decreases to typically ~[50,150] ms while approaching the NRL at the intermittent phase (**Figures 4** and **5A,D**). These two regions are also reflected in the behavior of $\sigma(L_{ST})$, where below a certain NRL, approximated by $L_T$, the standard deviation, $\sigma$, is almost constant and diverges as a power-law as the intermittent NRL is approached (**Figures 4** and **5C,F**). The maximal $\sigma$ varies much among neurons (**Figures 5B,E**), and similarly large deviations in the exponent characterizing the power-law are observed (**Figures 5C,F**). Nevertheless, the power-law was found to be a universal characteristic of the transition to the intermittent phase.

The NRL at the transition between the regions of semi-constant and rapidly increasing $\sigma$, $L_T$, is estimated where the NRL profile changes its concave form to convex, $d^2L/dStimulation^2=0$, under relatively high stimulation frequencies (**Figures 5A,D**). The concave profile, $L<L_T$, was recently found to identify a reproducible non-chaotic neuronal phase(Marmari et al., 2014). In this phase deviations between NRL profiles, obtained in different stimulation trials consisting of identical stimulation pattern of a single neuron, do not increase with the number of stimulations(Marmari et al., 2014). In an opposite manner, for a convex NRL profile, $L>L_T$, a chaotic neuronal phase emerges, indicating an exponential divergence among NRL profiles obtained in different trials. The emergence of non-chaotic and chaotic neuronal phases preceding the intermittent phase(Marmari et al., 2014) is also reflected in the NRL stabilization. Stabilization at $L_{ST}<L_T$ is realized with semi-constant small $\sigma$, whereas at $L_{ST}>L_T$ a rapid increase in $\sigma$ is observed.

The minimal standard deviation of few microseconds, i.e. $\sigma$~5 $\mu s$ (**Figures 5B,E**), is measured when $L_{ST}$ is close to the initial NRL, and requires sub-Hertz stimulation rate. These minimal deviations saturate our experimental lower bound of $\sigma$, which stems from the unavoidable amplified noise measured by the electrode (**Figure 6**), thus the examination of a better neuronal stabilization is beyond our experimental limitations. To overcome these limitations, we turn to the following theoretical argument. The number of ions in an evoked spike can be estimated from the density of the neuronal membrane capacitance which is around 1 $\mu F/cm^2$ (Cole, 1968;Gentet et al., 2000). Hence, for the diameter of a typical neuronal soma, ~20 $\mu m$ (its surface area ~$10^3$ $\mu m^2$)(Lübke et al., 1996), one finds the neuronal membrane capacitance as C~$10^{-11}$F. Using the voltage difference in the membrane during an evoked spike(Cole, 1968), $\Delta V$~0.1 V, the scaling of the number of ions in an evoked spike is estimated as Q=C$\Delta V$~$10^7$ e. These ~$10^7$ ions are evoked in about 1 ms, the duration of a spike, implying that an individual ion is evoked on the average every 1 ms/$10^7$=$10^{-10}$ s. Assuming a simple stochastic process for the sequential emission of ions, the expected time deviation for an evoked spike is $\Delta t \sim 10^{-10} \sqrt{10^7} \sim 0.3$ µs, which is only slightly below the experimentally demonstrated $\sigma$ of extremely few microseconds.

### 3.2.2 *In vivo* experiments

The NRL stabilization, with fluctuations in the order of 100 µs, was confirmed also in *in vivo* experiments (**Figure 7**). Results support the *in vitro* experiments such that $\sigma$ can reach ~100 µs without response failures despite 1-8 ms NRL stretching. However, in enhanced stimulation frequencies the intermittent phase emerged, characterized by increased fluctuations and the emergence of response failures. A precise characterization of the intermittent phase requires much longer stimulation periods, which are limited in our experimental setup, and also taking into account the neuronal spontaneous activity, which in the presented neuron was ~2 Hz. The ~1.5 mm distance between the stimulating and recording electrodes(Abeles, 1991;Kincaid et al., 1998;Zheng and Wilson, 2002) and the large (~17 ms) NRL stretching at the intermittent phase (**Figure 7D**) support a synaptic mechanism with high probability (through orthodromic stimulation) and probably even relaying via several synapses(Abeles, 1991;Kincaid et al., 1998;Zheng and Wilson, 2002). Note that the lack of a clear $L_T$ might be attributed to the accumulation of the NRL along a neuronal chain. These finding strongly support the temporal robustness of signal transmission in the brain(Csicsvari et al., 1998;Bonifazi et al., 2005;Boudkkazi et al., 2011), even in cases where both synaptic and neuronal transmission are involved.

### 3.3 Universal properties at the intermittent phase

Above a critical stimulation frequency, $f_C$, the neuron enters the intermittent phase, characterized by a maximal average NRL, $L_C$, independent of the stimulation frequency(Gal et al., 2010), accompanied by the appearance of response failures (**Figures 1A,8A**) and large $\sigma$ (**Figures 5B,E**). Specifically, there is a critical time-lag $\tau_C$ (corresponding to $f_C$) where $L_C$ is first achieved accompanied by large $\sigma$ but with a vanishing fraction of response failures (**Figure 8B**). The transient times to reach stable NRLs are comparable for $f \leq f_C$ (**Figures 1C, 4, 8A,B**). Nevertheless, as $f_C$ is approached the transient time to a stationary $\sigma$ is much enhanced (**Figure 8B**), together with a power-law divergence of $\sigma$ (**Figures 5C,F**), as expected in such a second order phase transition(Stanley, 1987).

At stimulation frequencies above $f_C$, neuronal response failures stochastically emerge such that their fraction is well approximated by $P_{fail}=(\tau_C-\tau)/\tau_C$ (**Figure 8C**). Consequently, a global tenable quantity, the average inter-spike-interval (ISI), is preserved and is equal to $\tau/(1-P_{fail})=\tau_C$ (**Figure**

**8C**), hence the neuron functions similar to a low pass filter. The stochastic occurrence of response failures was quantitatively examined using the following two tests. The first consists of calculating the probability for the occurrence of all combinations of responses for a given set of consecutive stimulations. These probabilities are then compared to the expected values under the assumption of random uncorrelated response failures with a given $P_{fail}$ (**Figure 8D**). The second test consists of calculating the probabilities, $P_0(m)$, for the occurrence of segments of m consecutive response failures bounded by evoked spikes. These probabilities were found to be in a good agreement with the theoretically predicted ones based on a Poissonian process with a rate - ln($P_{fail}$) (**Figure 8E**). Stimulation frequencies close to $f_C$ result in $P_{fail} << 0.5$, and enable the examination of stochastic response failures for much longer segments consisting of m evoked spikes bounded by response failures, $P_1(m)$ (**Figure 8F**). Both tests indicate that response failures emerge stochastically and independently at the intermittent phase.

Similar results for the neuronal critical frequency, $f_C$, as well as stabilization of the NRL were also found for cultured cortical neurons that were not functionally separated from their network by the addition of synaptic blockers (**Figure 9**), strengthening the biological relevance of our findings.

### 3.4 Fast neuronal adaptation to frequency modulation

The saturated neuronal firing frequency, $f_C$, functions as an impedance mechanism limiting the average neuronal firing rate. Typically, $f_C$ is in the range of 6-15 Hz, but can be extended for some neurons as high as 27 Hz and as low as 3 Hz.

A plausible biological scenario that suppresses the firing frequency of a single neuron below $f_C$ is aperiodic time-lags between stimulations, as verified experimentally (**Figure 10A**). For illustration, assume a slow mode of alternation between stimulation frequencies of $2f_C$ ($0.5\tau_C$ time-lag between stimulations) and $2f_C/3$ ($1.5\tau_C$), such that the average time-lag between stimulations is $\tau_C$. For the high and low frequency modes, the expected probability for response failures is 0.5 and 0, respectively. Consequently, the average ISI is $0.5(1.5\tau_C + \tau_C) = 1.25\tau_C$, corresponding to a lower firing rate of $0.8f_C$.

The fairly good agreement between the lowered firing rate, below $f_C$, and the predicted one (**Figure 10A**) strongly indicates fast neuronal plasticity (adaption) where the probability for a neuronal response failure is intrinsically adjusted following the temporary stimulation frequency. To quantify the time scale of this type of neuronal plasticity, a neuron characterized by $f_C \sim 5.5$ Hz was repeatedly stimulated with a recurrence of 80 stimulations, 40 at 12 Hz and 40 at 7 Hz (**Figure 10B**, inset). The probability for response failure for each of the 80 stimulations, measured over 200 recurrences, consists of two semi-stationary values; ~0.6 is attributed to stimulations given at 12 Hz and ~0.3 to stimulations given at 7 Hz (**Figure 10B**). The transient time between these values was found to vary between five and several dozen stimulations among neurons.

The probability profile of response failures, $P_{fail}$, (**Figure 10B**) was fitted to the function

$$P_{fail}(i) = A \sum_{m=i-80}^{i-1} \left( \frac{\tau_C - \tau_{(m)}}{\tau_C} \right) e^{-\alpha(i-m)} \qquad (3)$$

where the stimulation number $i$ is an integer in the range [0,79], $\tau_{(m)}$ is the time-lag between stimulations m and m+1, a negative m reads as $\tau_{(m+80)}$ and the term $(\tau_C - \tau_{(m)})/\tau_C$ represents the probability for a response failure following $\tau_{(m)}$. The last term represents the weighted exponential decay function with the optimized fitted fading coefficient $\alpha$ (**Figure 10B**) and $A$ is the normalization coefficient setting $P_{fail}(i)=1$ if all $\tau_{(m)}$ are zero (see MATERIALS AND METHODS).

This quantitative modeling of intrinsic short-term plasticity enables the examination of its abundant cooperative effects within neuronal chains and networks.

### 3.5 Neuronal impedance mechanism on a network level

We experimentally examined a chain of two neurons, characterized by the critical frequencies $f_C$=7.5 and 15.5 Hz for the first and second neuron, respectively (**Figure 10C**). The first neuron was stimulated with time-lags between stimulations taken randomly from $U(20,110)$ ms (**Figure 10C**), resulting in a ~7.3 Hz average firing rate, close to its $f_C$, and consequently its $\sigma$ is large, exceeding 350 μs (**Figure 10C**). The second neuron is stimulated by the first neuron at ~7.3 Hz on the average, far below its $f_C$=15.5 Hz. Hence, it relays the stimulations in the form of evoked spikes without response failures ($P_{fail}$~0) and with supreme precision, $\sigma$~40 μs (**Figure 10C**).

The effect of the neuronal response impedance mechanism on the network level was examined using large scale simulations of excitatory networks composed of N=2000 leaky integrate and fire neurons (see MATERIALS AND METHODS), whose prototypical topology was constructed using the following two steps. Each neuron is first randomly selected to have exactly one post-synaptic and one pre-synaptic connection, and the remaining connections are then selected with a survival probability of 0.1/N (**Figure 11A**). All connections are above-threshold, delays are taken randomly from $U(6, 9.5)$ ms and neurons are selected randomly to have either $f_C$~6.66 Hz ($\tau_C$=150 ms) or $f_C$~14.28 Hz ($\tau_C$=70 ms) (**Figure 11B**). Response failures for each neuron were implemented using the impedance mechanism, equation (1) with $\alpha$=1.4 (see MATERIALS AND METHODS), however results were found to be insensitive to the precise $\alpha$ (**Figure 12A**). The distribution of firing rates of the 2000 neurons was estimated after several seconds using a time window of about 50 s (**Figures 11B-E,12**), indicating ~5.4 Hz averaged firing rate (**Figure 11B**) which is even below the minimal $f_C$=6.66 Hz. This result is a consequence of the two abovementioned experimentally verified effects. The first, a single neuron along the chain, characterized by $f_C$=6.66 Hz, enforces all its consecutive neurons to fire no higher than this frequency. Since in a recurrent network most neurons have an ancestor neuron with such low $f_C$=6.66 Hz, they are expected to lower their firing rates towards this frequency. The second effect is aperiodic time-lags between stimulations, leading to firing rates even further below $f_C$=6.66 Hz (**Figure 11B**).

The cooperative effect of the neuronal response impedance mechanism drives the average firing rate of the entire network even below the lower tail of the distribution of $f_C$. In this state of low firing rates, neurons are typically in the non-chaotic phase, where deviations of only several microseconds around the average NRL are expected (**Figure 5**). Hence, low firing rates and microsecond neuronal precision are simultaneously achieved on a network level even in the presence of neurons which can potentially fire at very high rates(McCormick et al., 1985;Tateno et al., 2004). This tendency was found to be robust to a more realistic scenario where $f_C$ was taken randomly from $U$(6.66, 14.28) Hz (**Figure 11C**). Since our experiments indicate that $f_C$ can be as low as 3 Hz, compared to 6.66 Hz used in the above simulations, in large scale neural networks even lower firing rates are expected, as experimentally observed in cortical activity(Shafi et al., 2007;O'Connor et al., 2010).

The lack of response failures for stimulation frequencies below $f_C$ is too simplistic of an assumption for cortical dynamics, as failures may be generated, for instance, by synaptic noise and background inhibition. A theoretical argument and simulations (**Figure 11D**) indicate that indeed the proposed cooperative effect is robust to an additional response failure probability, $p<0.075$, for all frequencies, including those below $f_C$. This critical probability, 0.075, is a result of the average chain length, which for our network topology is ~9, as the probability for a neuron to have two post-synaptic connections is 0.1. Consequently, the probability for an evoked spike from the last neuron in the chain, given a stimulation to the first one is $(1-p)^9$. Since the chain terminates in a branch to two consecutive chains, the preference of spike birth over spike death requires $2(1-p)^9>1$, resulting in $p<0.075$. Similarly, low firing rates on a network level were found in simulations to be robust to the scenario of spontaneous stimulations at an average rate of 1 Hz per neuron (**Figure 11E**), as well as for the same network architecture with additional above- and sub-threshold connections, for both sparse and dense scenarios (**Figures 12B,C**).

## 4. DISCUSSION

The dynamical properties of networks are typically assumed to reflect the statistical properties of their links. Following this framework, the low firing rate for a given neural network topology was achieved in simulations using specific distributions of excitatory and inhibitory synapses, which balance spike birth and death trends. Since an inhibitory synapse, a directed link, probabilistically blocks an evoked spike of its driven node only in a given time window(Vardi et al., 2013b), the low firing rates are expected to be sensitive to small changes in network topology, synaptic delays and to the emergence of spontaneous activity(Daqing et al., 2011), unless a mean-field limit is assumed. In this work we experimentally present the neuronal response impedance mechanism on a single neuron level, which results in stochastic neuronal response failures at high stimulation frequencies and in precise response timings at low stimulation frequencies. On a network level, this mechanism leads to robust low firing rates, where each node, neuron, independently generates response failures above a critical stimulation frequency and functions similar to a low pass filter. Consequently, cooperation among individual neurons, enforced by the network dynamics, results in low firing rates which are governed by the low critical frequencies of the extreme nodes. As a byproduct of these low firing rates, the nodal response timings are stabilized with microsecond neuronal precision. The emergence of the two cooperative features on a network level supports the possibility that all building blocks of neural networks, neurons and synapses, jointly operate under the same extreme precision.

An indirect experimental support that low firing rate on the network level is achieved using solely neuronal response failures and without inhibition is also presented (**Figure 13**). We compare the spontaneous activity of the same MEA plated with cultured cortical neurons without additional synaptic blocker and with additional Bicuculline which blocks inhibition. In both cases, the average spontaneous activity of each electrode is measure over 10 minutes. Results, exemplified in **Figure 13**, clearly indicate that in both cases low firing rates are maintained, although the suppression of the inhibition slightly enhanced the spike detection rates of most of the electrodes. It is clear that typically each electrode records the spiking activity of more than one neuron, hence the firing activity recorded by the MEA does not directly count the activity of the entire neural network. Nevertheless, the only slight increase in average firing activity measure by the MEA, supports our findings that the phenomenon of low firing rates is mainly attributed to the neuronal response failures

The question arises as to what functionalities demand synaptic inhibition, inhibitory synapses. A possible hypothesis is that stationary network activity of low firing rates acts as a baseline cortical state. Over this state of activity, meaningful neuronal functionalities are embedded by the conditional temporal formation of neuronal firing, resulting in effective spatial summation, opening new routes of information flow through the network. Alternatively, temporary higher frequency stimulations to a subset of neurons result in abrupt changes of their NRLs (**Figure 3**). The accumulation of these changes along neuronal pathways(Vardi et al., 2013a;Vardi et al., 2013b;Goldental et al., 2014;Vardi et al., 2014) dynamically changes the topography of the network(Vardi et al., 2013b;Goldental et al., 2014). Both abovementioned scenarios only temporarily influence the baseline cortical state, as the NRL stretching is a fully reversible phenomenon(De Col et al., 2008;Vardi et al., 2013c).

   This work does not contradict the known mechanisms, e.g. inhibition, which lead to low firing rates. Nevertheless, the proposed mechanism is significantly robust to changes in the network structure, e.g. connectivity and synaptic strengths. Specifically, we have shown that neural network will exhibit low firing rates even without inhibition, as the proposed mechanism is based on the unreliable responses of the neurons. Our findings call for the reexamination of the role of inhibition as the main suppressor of firing rates in neural networks. Specifically, what is the synergism between inhibition and the intrinsic neuronal impedance mechanism and in what dynamical circumstances one is more dominant than the other? Furthermore we have shown that although neurons generally do not respond in a temporal precision and imprecisions accumulate along the network, the low firing rates lead to a supreme stabilization of the neuronal responses. This stabilization is characterized by imprecisions of only several μs, opening the doors for the feasibility of temporal coding in various functionalities in the brain.

# REFERENCES

Abeles, M. (1991). *Corticonics: Neural circuits of the cerebral cortex.* Cambridge University Press.

Agmon-Snir, H., Carr, C.E., and Rinzel, J. (1998). The role of dendrites in auditory coincidence detection. *Nature* 393**,** 268-272.

Amit, D.J., and Brunel, N. (1997). Model of global spontaneous activity and local structured activity during delay periods in the cerebral cortex. *Cerebral Cortex* 7**,** 237-252.

Bonifazi, P., Ruaro, M.E., and Torre, V. (2005). Statistical properties of information processing in neuronal networks. *European Journal of Neuroscience* 22**,** 2953-2964.

Boudkkazi, S., Fronzaroli-Molinieres, L., and Debanne, D. (2011). Presynaptic action potential waveform determines cortical synaptic latency. *The Journal of physiology* 589**,** 1117-1131.

Brama, H., Guberman, S., Abeles, M., Stern, E., and Kanter, I. (2014). Synchronization among neuronal pools without common inputs: in vivo study. *Brain Structure and Function***,** 1-11.

Brunel, N. (2000). Dynamics of sparsely connected networks of excitatory and inhibitory spiking neurons. *Journal of computational neuroscience* 8**,** 183-208.

Butts, D.A., Weng, C., Jin, J., Yeh, C.-I., Lesica, N.A., Alonso, J.-M., and Stanley, G.B. (2007). Temporal precision in the neural code and the timescales of natural vision. *Nature* 449**,** 92-95.

Carr, C.E. (1993). Processing of temporal information in the brain. *Annual review of neuroscience* 16**,** 223-243.

Chih, B., Engelman, H., and Scheiffele, P. (2005). Control of excitatory and inhibitory synapse formation by neuroligins. *Science* 307**,** 1324-1328.

Cole, K.S. (1968). *Membranes, ions, and impulses: a chapter of classical biophysics.* Univ of California Press.

Csicsvari, J., Hirase, H., Czurko, A., and Buzsáki, G. (1998). Reliability and state dependence of pyramidal cell–interneuron synapses in the hippocampus: an ensemble approach in the behaving rat. *Neuron* 21**,** 179-189.

Daqing, L., Kosmidis, K., Bunde, A., and Havlin, S. (2011). Dimension of spatially embedded networks. *Nature Physics* 7**,** 481-484.

De Col, R., Messlinger, K., and Carr, R.W. (2008). Conduction velocity is regulated by sodium channel inactivation in unmyelinated axons innervating the rat cranial meninges. *The Journal of Physiology* 586.

Diesmann, M., Gewaltig, M.-O., and Aertsen, A. (1999). Stable propagation of synchronous spiking in cortical neural networks. *Nature* 402**,** 529-533.

Doyle, M.W., and Andresen, M.C. (2001). Reliability of monosynaptic sensory transmission in brain stem neurons in vitro. *Journal of Neurophysiology* 85**,** 2213-2223.

Foust, A., Popovic, M., Zecevic, D., and Mccormick, D.A. (2010). Action potentials initiate in the axon initial segment and propagate through axon collaterals reliably in cerebellar Purkinje neurons. *The Journal of Neuroscience* 30**,** 6891-6902.

Gal, A., Eytan, D., Wallach, A., Sandler, M., Schiller, J., and Marom, S. (2010). Dynamics of excitability over extended timescales in cultured cortical neurons. *The Journal of Neuroscience* 30**,** 16332-16342.

Gentet, L.J., Stuart, G.J., and Clements, J.D. (2000). Direct measurement of specific membrane capacitance in neurons. *Biophysical Journal* 79**,** 314-320.

Goldental, A., Guberman, S., Vardi, R., and Kanter, I. (2014). A computational paradigm for dynamic logic-gates in neuronal activity. *Frontiers in computational neuroscience* 8.

He, B.J., Zempel, J.M., Snyder, A.Z., and Raichle, M.E. (2010). The temporal structures and functional significance of scale-free brain activity. *Neuron* 66**,** 353-369.

Kayser, C., Logothetis, N.K., and Panzeri, S. (2010). Millisecond encoding precision of auditory cortex neurons. *Proceedings of the National Academy of Sciences* 107**,** 16976-16981.

Kincaid, A.E., Zheng, T., and Wilson, C.J. (1998). Connectivity and convergence of single corticostriatal axons. *The Journal of neuroscience* 18**,** 4722-4731.

Kumar, A., Rotter, S., and Aertsen, A. (2010). Spiking activity propagation in neuronal networks: reconciling different perspectives on neural coding. *Nature Reviews Neuroscience* 11**,** 615-627.

Lass, Y., and Abeles, M. (1975). Transmission of information by the axon: I. Noise and memory in the myelinated nerve fiber of the frog. *Biological Cybernetics* 19**,** 61-67.

Litvak, V., Sompolinsky, H., Segev, I., and Abeles, M. (2003). On the transmission of rate code in long feedforward networks with excitatory–inhibitory balance. *The Journal of neuroscience* 23**,** 3006-3015.

Lübke, J., Markram, H., Frotscher, M., and Sakmann, B. (1996). Frequency and dendritic distribution of autapses established by layer 5 pyramidal neurons in the developing rat neocortex: comparison with synaptic innervation of adjacent neurons of the same class. *The Journal of neuroscience* 16**,** 3209-3218.

Mainen, Z.F., and Sejnowski, T.J. (1995). Reliability of spike timing in neocortical neurons. *Science* 268**,** 1503-1506.

Marmari, H., Vardi, R., and Kanter, I. (2014). Chaotic and non-chaotic phases in experimental responses of a single neuron. *EPL (Europhysics Letters)* 106**,** 46002.

Mccormick, D.A., Connors, B.W., Lighthall, J.W., and Prince, D.A. (1985). Comparative electrophysiology of pyramidal and sparsely spiny stellate neurons of the neocortex. *J Neurophysiol* 54**,** 782-806.

O'connor, D.H., Peron, S.P., Huber, D., and Svoboda, K. (2010). Neural activity in barrel cortex underlying vibrissa-based object localization in mice. *Neuron* 67**,** 1048-1061.

Panzeri, S., Brunel, N., Logothetis, N.K., and Kayser, C. (2010). Sensory neural codes using multiplexed temporal scales. *Trends in neurosciences* 33**,** 111-120.

Rad, A.A., Sendiña-Nadal, I., Papo, D., Zanin, M., Buldu, J.M., Del Pozo, F., and Boccaletti, S. (2012). Topological measure locating the effective crossover between segregation and integration in a modular network. *Physical review letters* 108**,** 228701.

Rodríguez-Moreno, A., Kohl, M.M., Reeve, J.E., Eaton, T.R., Collins, H.A., Anderson, H.L., and Paulsen, O. (2011). Presynaptic induction and expression of timing-dependent long-term depression demonstrated by compartment-specific photorelease of a use-dependent NMDA receptor antagonist. *The Journal of Neuroscience* 31**,** 8564-8569.

Schoppa, N., and Westbrook, G. (1999). Regulation of synaptic timing in the olfactory bulb by an A-type potassium current. *Nature neuroscience* 2**,** 1106-1113.

Shafi, M., Zhou, Y., Quintana, J., Chow, C., Fuster, J., and Bodner, M. (2007). Variability in neuronal activity in primate cortex during working memory tasks. *Neuroscience* 146**,** 1082-1108.

Song, S., Miller, K.D., and Abbott, L.F. (2000). Competitive Hebbian learning through spike-timing-dependent synaptic plasticity. *Nature neuroscience* 3**,** 919-926.

Spiegel, I., Mardinly, A.R., Gabel, H.W., Bazinet, J.E., Couch, C.H., Tzeng, C.P., Harmin, D.A., and Greenberg, M.E. (2014). Npas4 Regulates Excitatory-Inhibitory Balance within Neural Circuits through Cell-Type-Specific Gene Programs. *Cell* 157**,** 1216-1229.

Stanley, H.E. (1987). Introduction to phase transitions and critical phenomena. *Introduction to Phase Transitions and Critical Phenomena, by H Eugene Stanley, pp. 336. Foreword by H*

*Eugene Stanley. Oxford University Press, Jul 1987. ISBN-10: 0195053168. ISBN-13: 9780195053166* 1.

Steingrube, S., Timme, M., Wörgötter, F., and Manoonpong, P. (2010). Self-organized adaptation of a simple neural circuit enables complex robot behaviour. *Nature Physics* 6**,** 224-230.

Stern, E.A., Bacskai, B.J., Hickey, G.A., Attenello, F.J., Lombardo, J.A., and Hyman, B.T. (2004). Cortical synaptic integration in vivo is disrupted by amyloid-β plaques. *The Journal of neuroscience* 24**,** 4535-4540.

Stern, E.A., Kincaid, A.E., and Wilson, C.J. (1997). Spontaneous subthreshold membrane potential fluctuations and action potential variability of rat corticostriatal and striatal neurons in vivo. *Journal of Neurophysiology* 77**,** 1697-1715.

Tateno, T., Harsch, A., and Robinson, H. (2004). Threshold firing frequency–current relationships of neurons in rat somatosensory cortex: type 1 and type 2 dynamics. *Journal of neurophysiology* 92**,** 2283-2294.

Teramae, J.-N., Tsubo, Y., and Fukai, T. (2012). Optimal spike-based communication in excitable networks with strong-sparse and weak-dense links. *Scientific reports* 2.

Turrigiano, G.G. (2008). The self-tuning neuron: synaptic scaling of excitatory synapses. *Cell* 135**,** 422-435.

Turrigiano, G.G., and Nelson, S.B. (2004). Homeostatic plasticity in the developing nervous system. *Nature Reviews Neuroscience* 5**,** 97-107.

Van Vreeswijk, C., and Sompolinsky, H. (1998). Chaotic balanced state in a model of cortical circuits. *Neural computation* 10**,** 1321-1371.

Vanrullen, R., Guyonneau, R., and Thorpe, S.J. (2005). Spike times make sense. *Trends in neurosciences* 28**,** 1-4.

Vardi, R., Goldental, A., Guberman, S., Kalmanovich, A., Marmari, H., and Kanter, I. (2013a). Sudden synchrony leaps accompanied by frequency multiplications in neuronal activity. *Frontiers in Neural Circuits* 7.

Vardi, R., Guberman, S., Goldental, A., and Kanter, I. (2013b). An experimental evidence-based computational paradigm for new logic-gates in neuronal activity. *Europhysics Letters* 103**,** 66001.

Vardi, R., Marmari, H., and Kanter, I. (2014). Error correction and fast detectors implemented by ultrafast neuronal plasticity. *Physical Review E* 89**,** 042712.

Vardi, R., Timor, R., Marom, S., Abeles, M., and Kanter, I. (2012). Synchronization with mismatched synaptic delays: A unique role of elastic neuronal latency. *Europhysics Letters* 100**,** 48003.

Vardi, R., Timor, R., Marom, S., Abeles, M., and Kanter, I. (2013c). Synchronization by elastic neuronal latencies. *Physical Review E* 87**,** 012724.

Vogels, T., Sprekeler, H., Zenke, F., Clopath, C., and Gerstner, W. (2011). Inhibitory plasticity balances excitation and inhibition in sensory pathways and memory networks. *Science* 334**,** 1569-1573.

Vogels, T.P., and Abbott, L. (2009). Gating multiple signals through detailed balance of excitation and inhibition in spiking networks. *Nature neuroscience* 12**,** 483-491.

Wagenaar, D.A., Pine, J., and Potter, S.M. (2004). Effective parameters for stimulation of dissociated cultures using multi-electrode arrays. *Journal of neuroscience methods* 138**,** 27-37.

Zheng, T., and Wilson, C. (2002). Corticostriatal combinatorics: the implications of corticostriatal axonal arborizations. *Journal of neurophysiology* 87**,** 1007-1017.

## ACKNOWLEDGEMENTS

We thank Moshe Abeles for stimulating discussions. Invaluable computational assistance by Igor Reidler and technical assistance by Hana Arnon are acknowledged. This research was supported by the Ministry of Science and Technology, Israel.

## AUTHOR CONTRIBUTION

R.V. and H.M. prepared the cultural tissues, the experimental materials and performed the in vitro experiments. A.G. performed the simulations and participated in the development of the theoretical arguments. R.V., H.M. and A.G. analysed the data. S.S. participated in some of the in vitro experiments. P.S. designed the interface for the real-time experiments. H.B., E.S. and I.K. designed, performed and analyzed the in vivo experiments. R.V. and A.G. prepared the manuscript. I.K. initiated the study and supervised all aspects of the work. All authors discussed the results and commented on the manuscript.

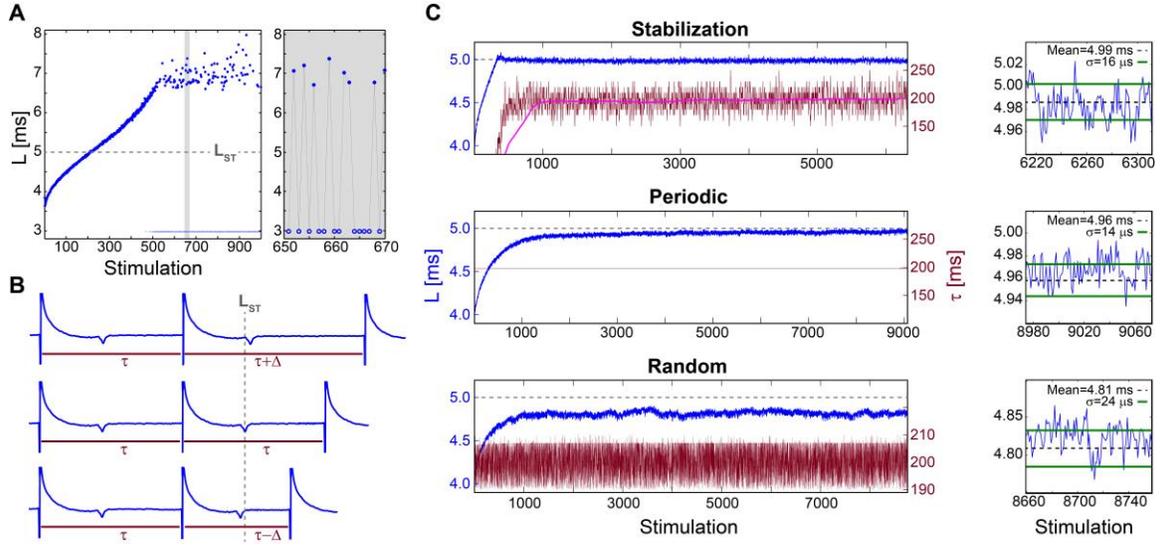

**FIGURE 1 | Stabilization of the neuronal response latency. (A)** The NRL, L, (blue dots) of a cultured neuron stimulated at 30 Hz, with a guideline for stabilization at $L_{ST}$=5 ms (gray dashed-line). Response failures (blue circles) are denoted at L=3 ms, exemplified by a zoom-in (gray). A guideline for stabilization at $L_{ST}$=5 ms is shown (gray dashed-line). **(B)** Schematic of a real-time adaptive algorithm for stabilization at $L_{ST}$. L(i) stands for the NRL to the $i^{th}$ stimulation, and τ(i) stands for the time-lag between the consecutive stimulations i-1 and i. In the event L(i)<$L_{ST}$ the next time-lag between stimulations is shortened, τ(i+1)=τ(i)-Δ (lower panel), whereas for L(i)>$L_{ST}$ it is enlarged, τ(i+1)=τ(i)+Δ (upper panel), otherwise it remains unchanged (middle panel). The step Δ is a predefined constant, which in advanced algorithms can be adjusted following the history of deviations from $L_{ST}$ (not shown). **(C)** Stabilization of L (blue) at $L_{ST}$~5 ms (gray dashed-line, also shown in **(A)**) and the time-lag τ between stimulations (crimson) for: the adaptive algorithm (described in **(B)**) using Δ=20 ms, where the smoothed τ using 1000 Stimulation sliding window (pink) saturates at ~198 ms (upper panel), periodic stimulation with τ=198.2 ms (middle panel) and τ taken randomly from $U$(190, 210) ms (lower panel). A zoom-in of L at the last 100 stimulations (blue) and the averaged L (black dashed-line) together with the standard deviation σ (green lines) obtained from the last 1000 stimulations for each stimulation scenario (right). All methods indicate a supreme stabilization, σ/L<$10^{-2}$. All experiments shown in this figure were done on a cultured neuron.

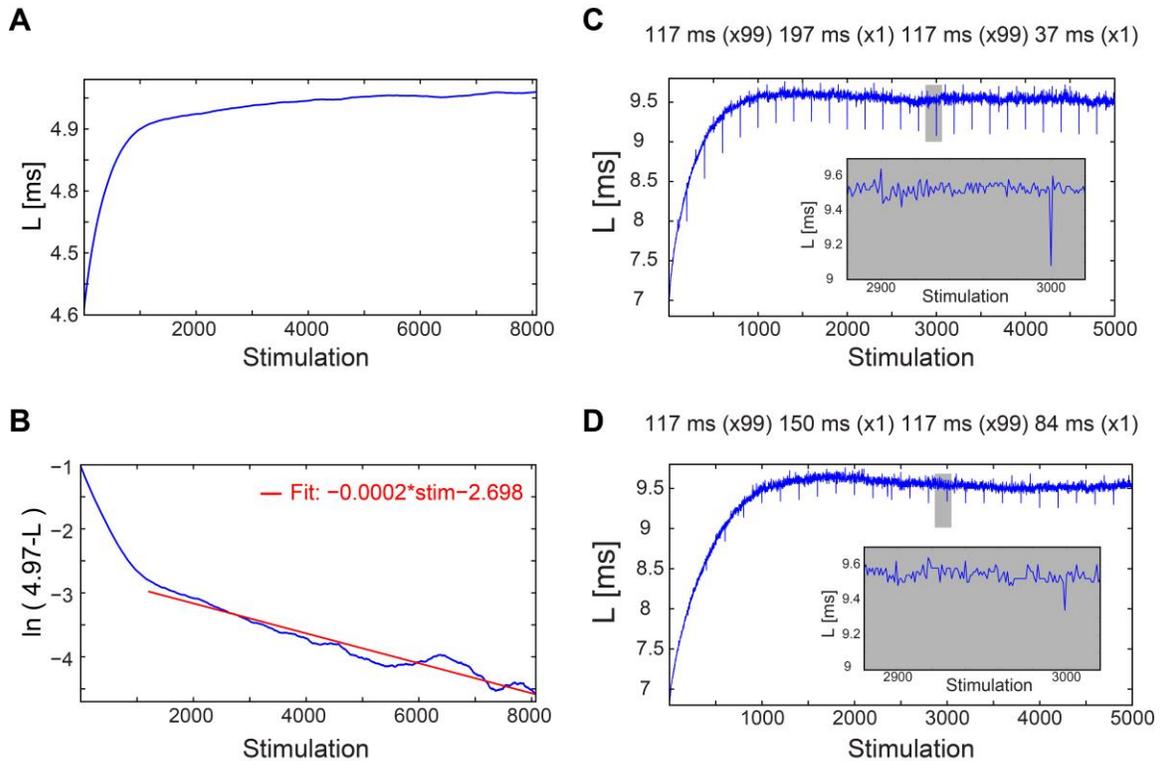

**FIGURE 2 | Convergence of the neuronal response latency to stabilization without adaptive external feedback, and asymmetry between facilitation and depression. (A)** The NRL, L, of a cultured neuron stimulated at $\tau$=198.2 ms between stimulations (blue), with an estimated $L_{ST}$ of 4.97 ms (**Figure 1C**, middle panel), smoothed using 1000 Stimulation sliding window. **(B)** $\ln(L_{ST}-L)$ using the estimated $L_{ST}$ (blue). The fit in the range of [1200, 8073] stimulations (red) indicates a convergence to $L_{ST}$ as $L_{ST}-L\sim0.06\cdot\exp(-0.0002\cdot$Stimulation). **(C)** The NRL, L, of a neuron stimulated at $\tau$=117 ms between stimulations, where at stimulations $(2n+1)\cdot100$ $\tau$=197 ms and at $(2n+2)\cdot100$ $\tau$=37 ms, where n=0,1,.... Results indicate an asymmetry between facilitation, with an amplitude of ~0.4 ms in case $\tau$ decreased, and depression, with an amplitude of ~0.1 ms in case $\tau$ increased (see zoom-in, gray). **(D)** Same as **(C)** but with moderate changes in $\tau$, at stimulations $(2n+1)\cdot100$ $\tau$=150 ms and at $(2n+2)\cdot100$ $\tau$=84 ms. Results indicate an asymmetry between facilitation, with an amplitude of ~0.2 ms in case $\tau$ decreased, and depression, with an amplitude of ~0.1 ms in case $\tau$ increased (see zoom-in, gray).

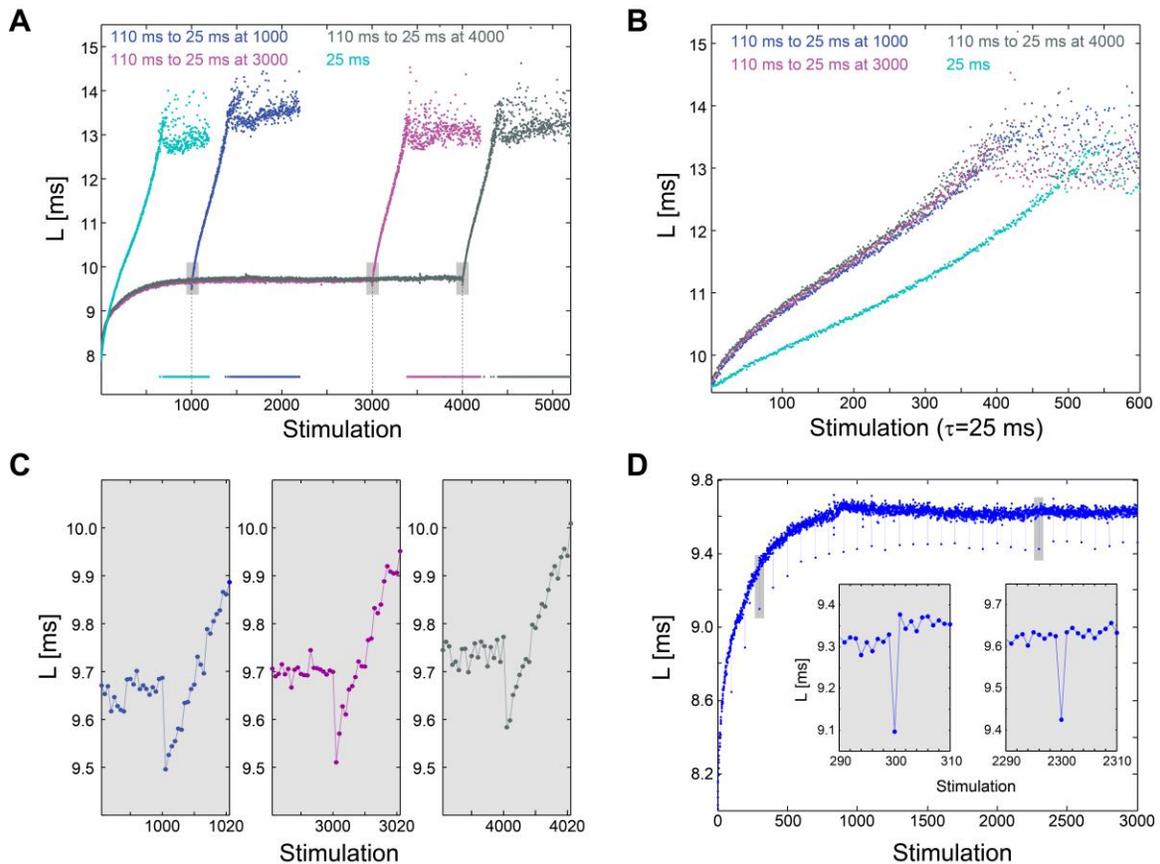

**FIGURE 3 | Neuronal plasticity at the stabilized neuronal response latency phase. (A)** Stabilization of the neuronal response latency, L, of a cultured neuron at $L_{ST}$~9.7 ms using $\tau$=110 ms time-lags between stimulations, followed by an abrupt change to $\tau$=25 ms, from stimulation 1000 (blue), 3000 (purple) or 4000 (dark gray). Response failures are denoted at L=7.5 ms. For comparison, L for a periodic stimulation with $\tau$=25 ms is shown (cyan). **(B)** Segments of the neuronal response latency profiles following the abrupt change to $\tau$=25 ms (shown in **(A)**), and the profile for L>$L_{ST}$~9.7 ms for periodic stimulation with $\tau$=25 ms. **(C)** A zoom-in of the transitions from $\tau$=110 to 25 ms (shown by gray boxes in **(A)**). **(D)** The same neuron as in **(A)**, where $\tau$ is shortened from 110 ms to 25 ms once every 100 stimulations. Two zoom-ins (gray boxes), indicate momentary facilitation with an amplitude ~0.2 ms.

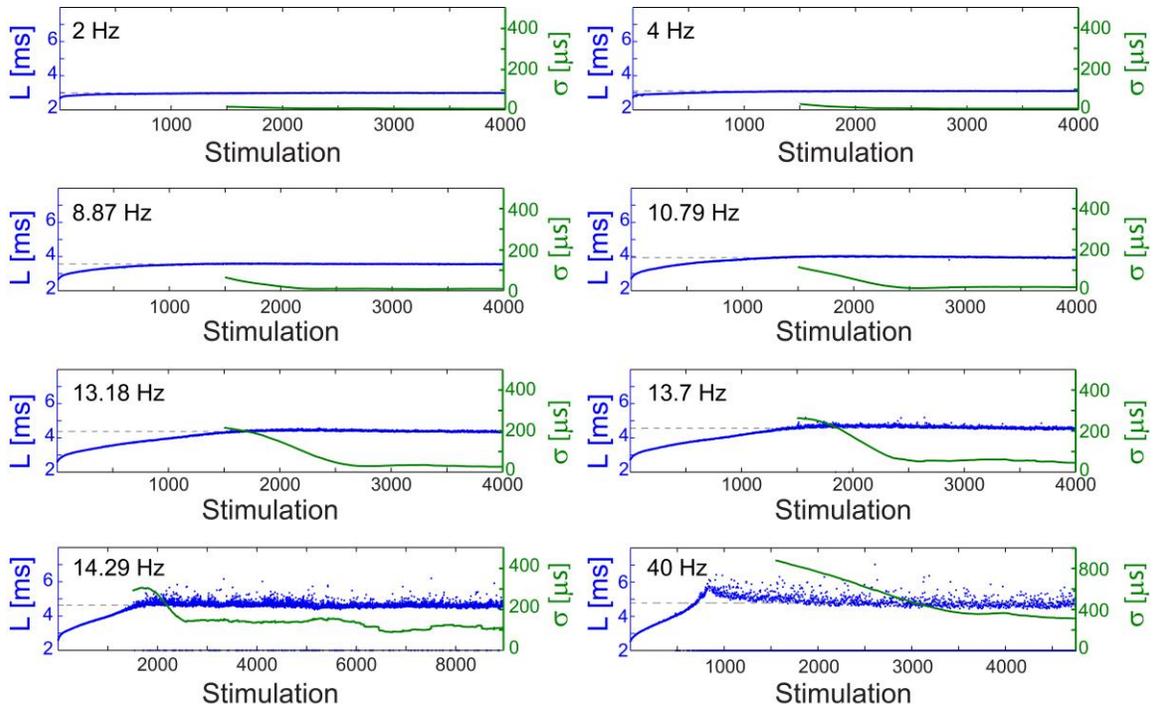

**FIGURE 4 | Stabilization at different latencies.** The NRL, L, of a cultured neuron under periodic stimulations at 2 Hz (500 ms), 4 Hz (250 ms), 8.87 Hz (112.74 ms), 10.79 Hz (92.68 ms), 13.18 Hz (75.87 ms), 13.7 Hz (72.99 ms), 14.29 Hz (69.98 ms) and 40 Hz (25 ms) (blue). The stabilization latencies are marked by the dashed-lines (average L over the last 500 responses). Response failures are denoted at L=2 ms. The standard deviation, σ, is shown smoothed using 1000 responses sliding window (green).

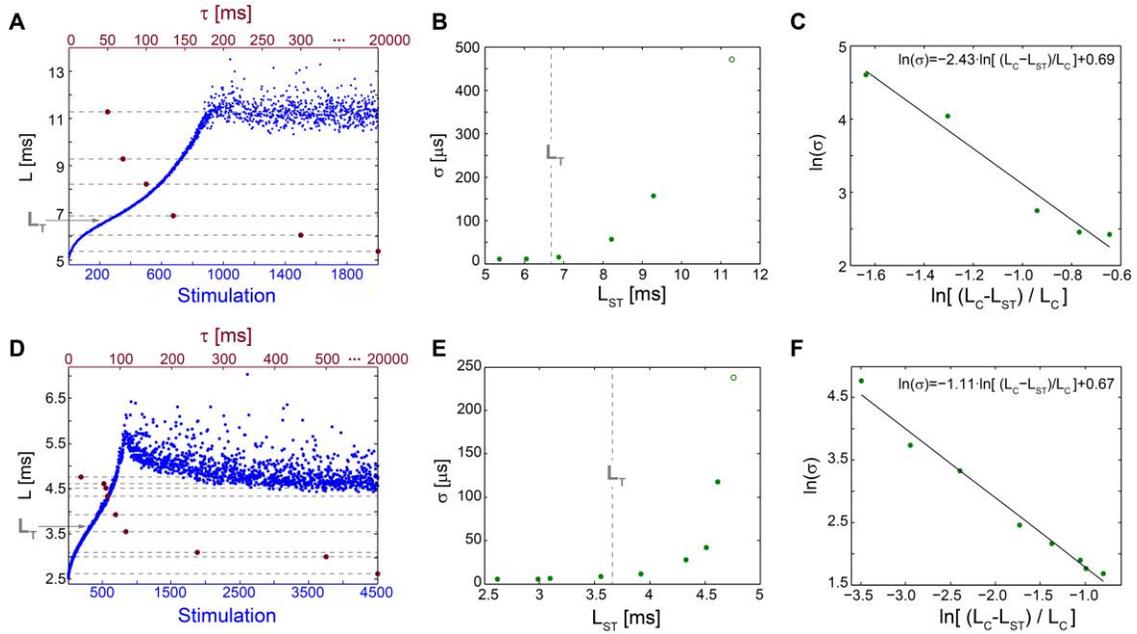

**FIGURE 5 | Neuronal temporal precision in the chaotic and non-chaotic phases. (A)** The NRL, L, of a cultured neuron stimulated at 20 Hz (blue), response failures are not shown, and stabilization latencies (dashed-lines) achieved by periodic stimulations characterized by various $\tau$ (crimson dots). **(B)** Standard deviations, $\sigma$, obtained from the last 1000 Stimulations for each of the stabilization latencies, $L_{ST}$, shown in **(A)** (green dots) and for the latency at the intermittent phase, $L_C$=11.28 ms (green circle). $L_T$ is shown as a guideline (vertical dashed-line). **(C)** A linear fit (black line) for $\ln(\sigma)$ (green dots) versus $\ln[(L_C-L_{ST})/L_C]$. **(D-F)** The same as **(A-C)** for a different cultured neuron stimulated at 40 Hz, with $L_T$=3.66 ms and $L_C$=4.73 ms.

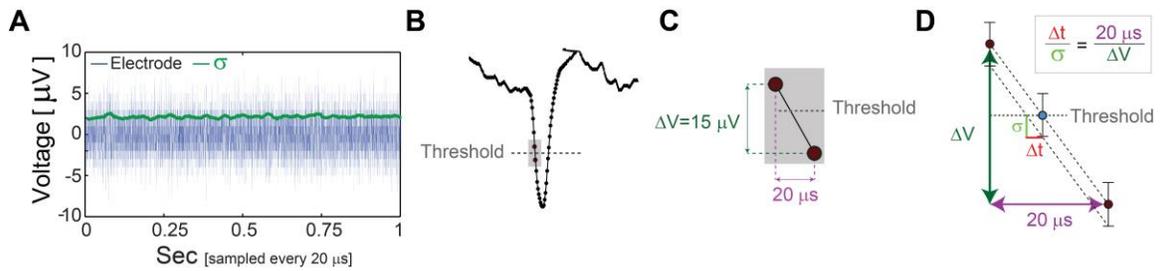

**FIGURE 6 | The unavoidable amplified noise measured by the MEA electrode. (A)** The amplified noise measured by one of the MEA electrode (during an *in vitro* experiment) over several minutes without stimulations (one second is shown, blue), sampled at 50 kHz (20 µs). The standard deviations, σ, smoothed using 1000 samples sliding window (green) indicate an average of 2.14 µV. **(B)** Schematic of a recorded spike, where dots represent voltage samples and the threshold for spike detection (horizontal gray dashed-line) is crossed between two consecutive samples. **(C)** A zoom-in of the two consecutive samples before and after the threshold crossing. **(D)** The linear interpolation of the time for threshold crossing (blue circle). The electrode-amplified noise, σ, is presented by the horizontal bars around the blue circle. This uncertainty in the electrode voltage induces an uncertainty at the interpolated threshold crossing time, Δt, following the relation $\frac{\Delta t}{\sigma} = \frac{20}{\Delta V}$. The time-lag between two samples is 20 µs (50 kHz sampling rate), and the average voltage difference between two samples in this spike phase is ~15 µV. For the above-mentioned typical values, σ=2.14 µV and ΔV=15 µV, one finds Δt=2.85 µs.

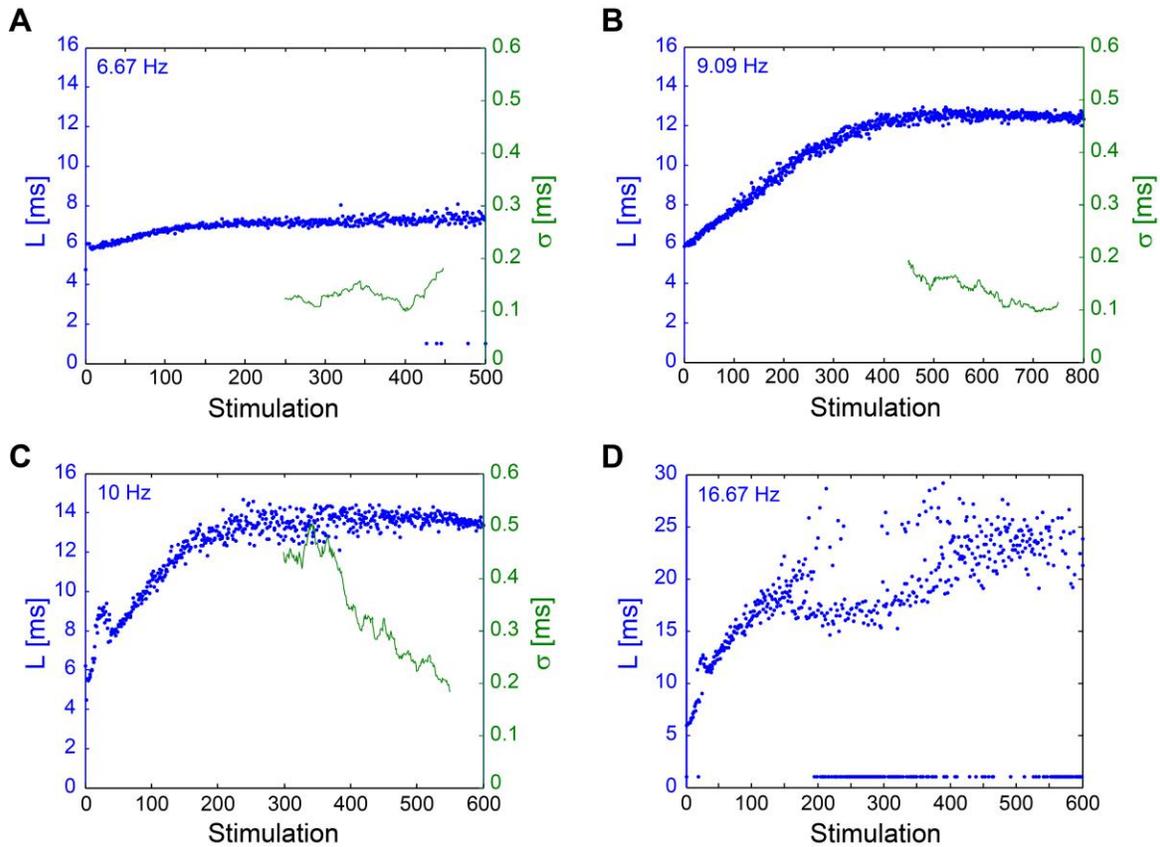

**FIGURE 7 | NRL stabilization measured using an *In Vivo* experimental setup.** Examples of the NRL, L, (blue) of a neuron recorded intracellularly *in vivo*, where stimulations were given extracellularly, ~1.5 mm away, (see MATERIALS AND METHODS, *in vivo* experiments section, similar to previous publications(Stern et al., 1997;Stern et al., 2004;Brama et al., 2014)), at various stimulation frequencies. Response failures are denoted at L~1 ms. The standard deviation ($\sigma$, green), smoothed using a 50 Stimulation window, is shown for **(A-C)**. **(A)** Stimulations at 6.67 Hz. The NRL stabilizes at $L_C$~7.2 ms, while $\sigma$ is relatively constant around 0.1 ms. **(B)** Stimulations at 9.09 Hz. The NRL stabilizes at $L_C$~12.5 ms, while $\sigma$ is stabilized around 0.1 ms. **(C)** Stimulations at 10 Hz. The NRL stabilizes at $L_C$~13.6 ms, while $\sigma$ decreases towards and even below 0.2 ms. **(D)** Stimulations at 16.67 Hz. The NRL at the intermittent phase fluctuates around $L_C$=23 ms, indicating ~17 ms NRL stretching. A relatively large amount of response failures emerge, such that $f_C$~15 Hz.

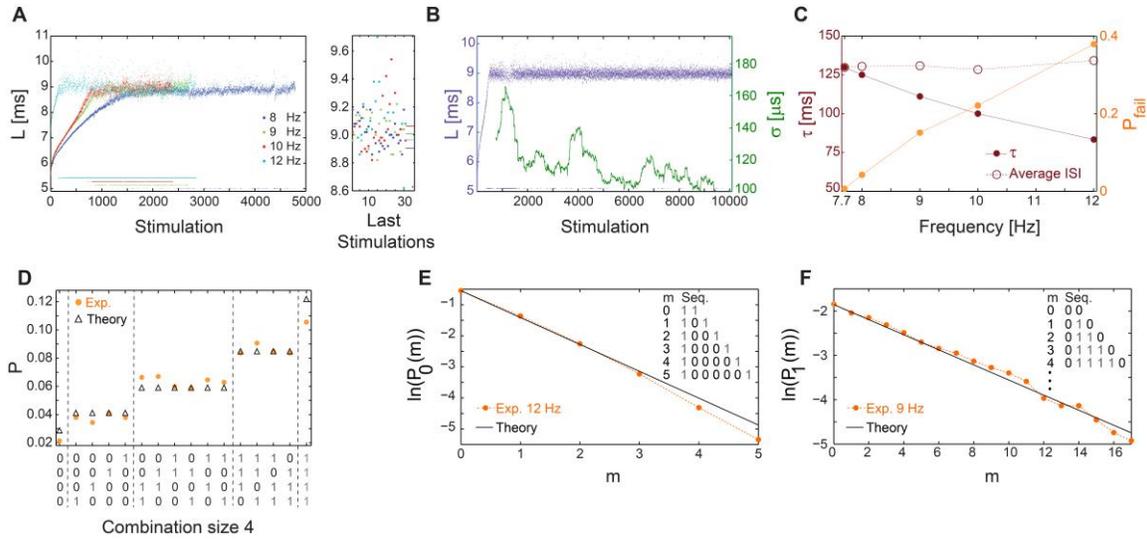

**FIGURE 8 | Universal properties of the stochastic response failures at the intermittent phase. (A)** The NRL, L, of a cultured neuron stimulated at 8 (blue), 9 (green), 10 (red) and 12 (cyan) Hz, response failures are denoted below L=6 ms. A zoom-in of L (right) obtained from the last 35 stimulations, response failures are not shown, with corresponding averaged L over the last 1000 stimulations resulting in 8.905, 9.007, 9.062, 8.964 ms for 8, 9, 10, 12 Hz, respectively (right horizontal bars), indicates $L_C \sim 9$ ms. **(B)** The neuron shown in **(A)** stabilized using the real-time adaptive algorithm (**Figure 1B**) with $L_{ST}=9$ ms. Response failures are denoted at $L \sim 5$ ms. The standard deviation, σ, smoothed using 500 Stimulation sliding window (green). **(C)** The probability for response failures, $P_{fail}$ (orange dots) and the corresponding τ (crimson dots) for each stimulation frequency in **(A)** and 7.7 Hz in **(B)**. The averaged ISI=$\tau/(1-P_{fail})$ (crimson circles). **(D)** The probabilities for all possible neuronal responses to four consecutive stimulations (x-axis) for the neuron in **(C)** stimulated at 12 Hz (orange dots), where 1/0 stand for evoked spike/response failure. These probabilities are compared to the corresponding theoretical probabilities (black triangles) assuming uncorrelated failures with the measured $P_{fail} \sim 0.4$. All measurements were taken at the intermittent phase. **(E)** The probabilities for the occurrence of segments of m consecutive response failures bounded by evoked spikes, $P_0(m)$, measured at the intermittent phase for the neuron in **(C)** stimulated at 12 Hz (orange dots). The theoretical values, $\ln(1-P_{fail})+m \cdot \ln(P_{fail})$ with $P_{fail}=0.4$ (black line), similar to **(C)**. **(F)** Similar to **(E)**, the probabilities for the occurrence of segments of m consecutive evoked spikes bounded by response failures, $P_1(m)$, measured at the intermittent phase for the neuron in **(C)** stimulated at 9 Hz (orange dots). The theoretical values, $\ln(P_{fail})+m \cdot \ln(1-P_{fail})$ with $P_{fail}=0.15$ (black line), similar to **(C)**. All experiments shown in this figure were done on a cultured neuron.

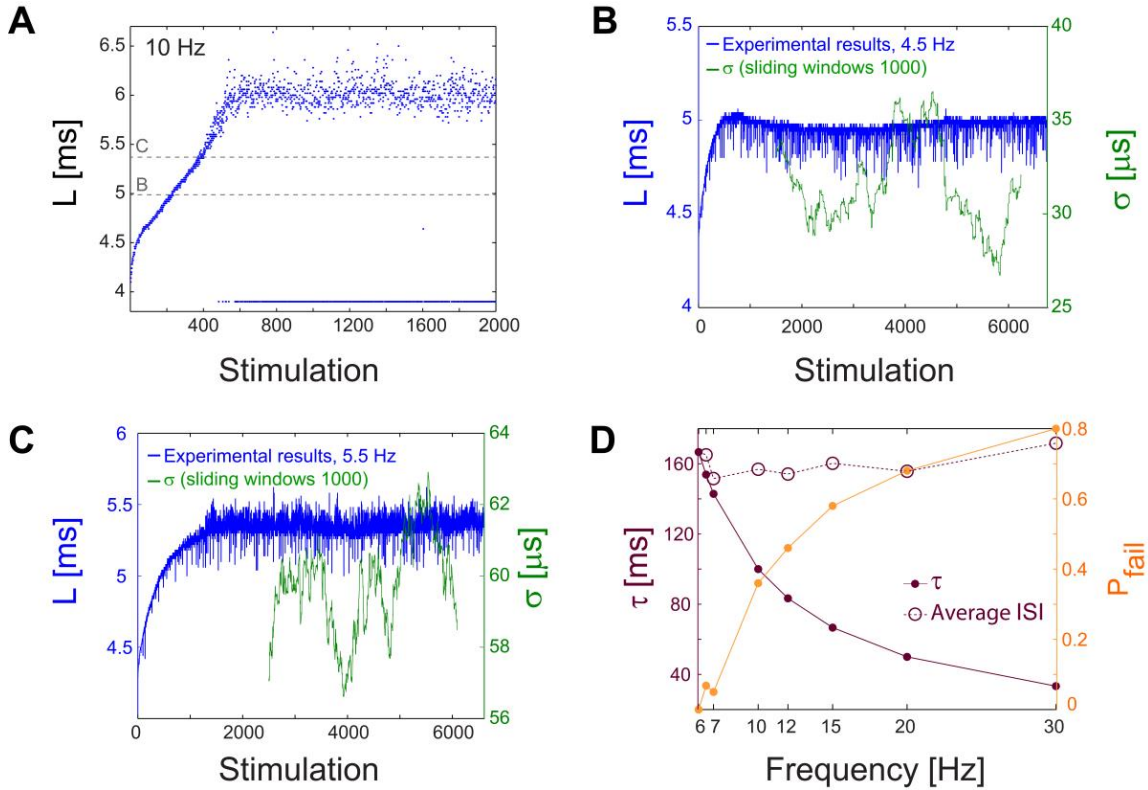

**FIGURE 9 | NRL measured in unblocked cultures of cortical neurons.** (**A**) The NRL, L, of a neuron embedded within a large-scale network of cortical cells *in vitro* (but not functionally separated from the network by synaptic blockers, see MATERIALS AND METHODS), stimulated at 10 Hz. At the intermittent phase response failures occur (denoted as dots at L~4 ms), and the NRL stabilizes at $L_C$~6.4 ms. The latencies discussed in the following panels, in (**B**) at L~5 ms, and in (**C**) at L~5.45 ms, are shown as guidelines (dashed gray lines). (**B**) The NRL, L, for a stimulation frequency of 4.5 Hz (blue). The average NRL over the last 2000 responses is ~4.9 ms. The standard deviation, $\sigma$, is shown smoothed using 1000 Stimulation sliding window (green). (**C**) Same as (**B**) but for a stimulation frequency of 5.5 Hz, resulting in an average NRL of L~5.4 ms measured over the last 2000 latencies. (**D**) The experimentally measured probability for response failures, $P_{fail}$ (orange dots) and the corresponding $\tau$ (crimson dots) at different frequencies in the range of [6, 30] Hz. For each frequency the averaged ISI=$\tau/(1-P_{fail})$ (crimson circles) is very close to $\tau_C$=160 ms ($f_C$~6.2 Hz), independent of the stimulation frequency.

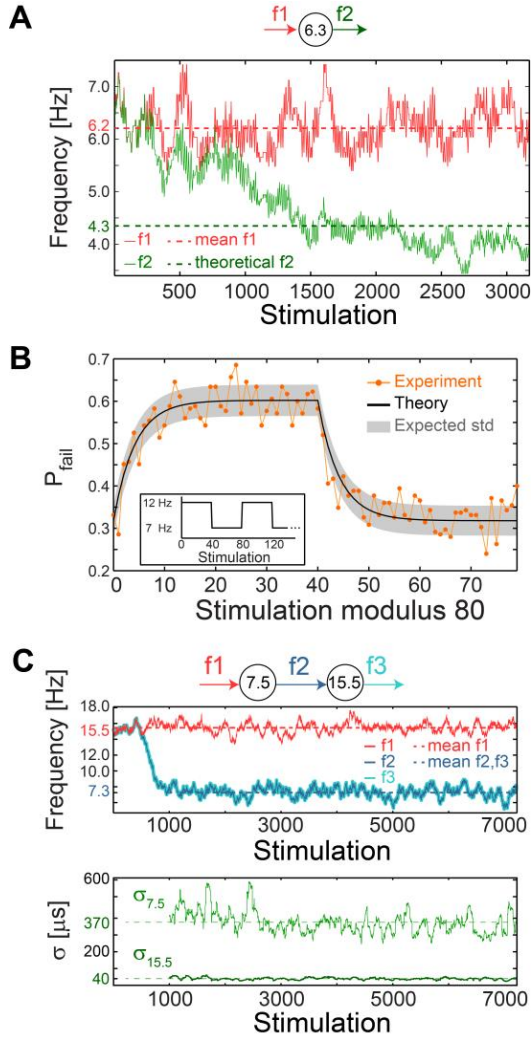

**FIGURE 10 | Neuronal impedance mechanism on a chain level leading to low firing rates. (A)** Schematic (top) of a cultured neuron with $f_C$=6.3 Hz and its stimulation/firing frequencies, f1/f2, respectively. Stimulation frequency (red) and firing frequency (green) smoothed using 100 Stimulation sliding window of a neuron stimulated with alternating time-lags, 20 and 300 ms, equal on the average to ~$\tau_C$, each time-lag repeats 5 to 10 times before switching. The averaged f1=6.2 Hz (red dashed-line) and the theoretically predicted f2=4.3 ms (green dashed-line) are shown for comparison. The resulting firing frequency of the cultured neuron is ~4 Hz (green full line), substantially lower than its $f_C$, and is close to the predicted f2 following ISI=0.5(160+300)=230 ms. **(B)** A cultured neuron with $f_C$~5.5 Hz was stimulated periodically, 40 times at 12 Hz and 40 times at 7 Hz (inset). The probability of a response failure, $P_{fail}$(i), i=0, …,79 (orange dots), measured over 200 recurrences, fitted with optimized $\alpha$=1.1 following equation (3) (black line) and the expected standard deviation $\sqrt{\left(P_{fail}(i)\left(1-P_{fail}(i)\right)\right)/200}$ (gray area). The experimentally observed probabilities are similar to the predicted values under a fixed stimulation frequency, $P_{fail}$=($\tau_C$-$\tau$)/$\tau_C$=1-$f_C$/f, resulting in 1-(5.5/12)~0.54 for 12 Hz and 1-(5.5/7)~0.21 for 7 Hz. **(C)** Schematic (top) of a chain of two cultured neurons with $f_C$=7.5 and 15.5 Hz, their stimulation and firing frequencies. Top panel: the firing frequency of the first/second neuron (dark/light blue), obtained when the first neuron is stimulated with $\tau$ taken randomly from $U$(20, 110) ms (red full-line), 15.5 Hz on the average (red dashed-line). This results in f2=f3 with an average of ~7.3 Hz (blue dashed-line). Bottom panel: $\sigma$ for the first/second neuron (light/dark green). Curves were smoothed using 100 Stimulation sliding window.

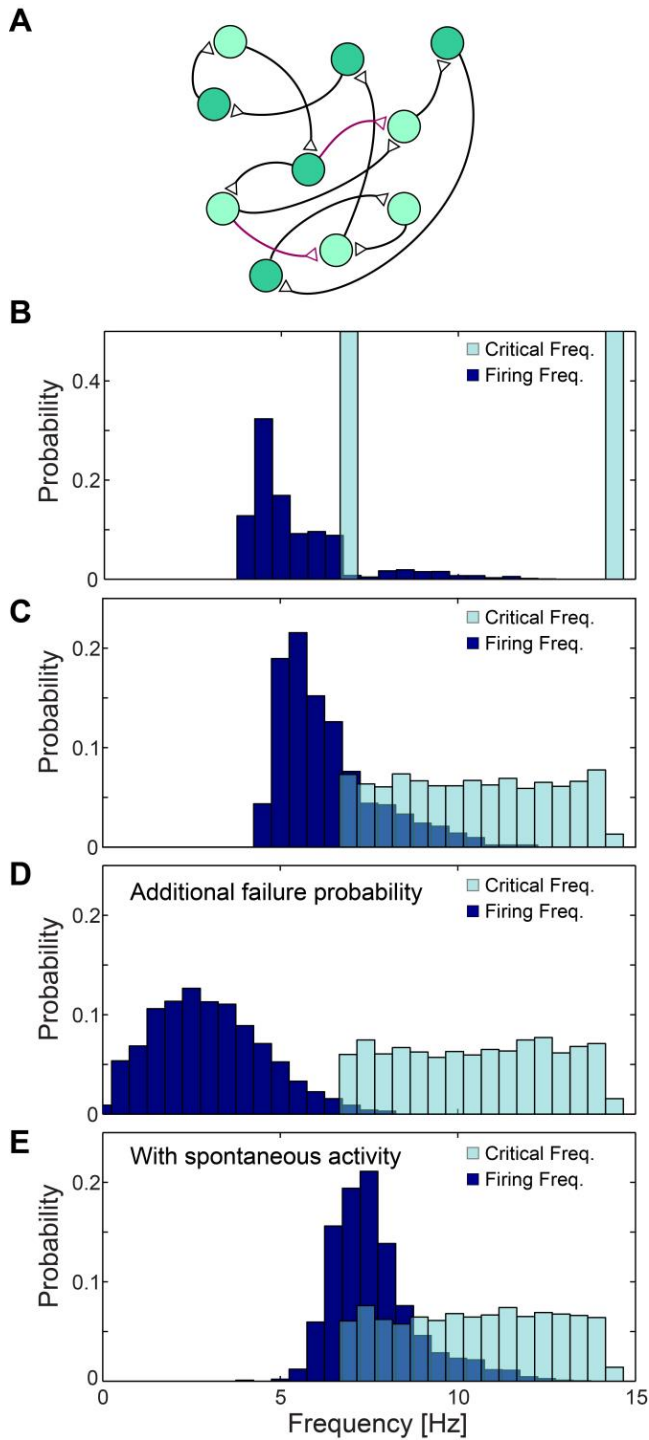

**FIGURE 11 | Neuronal cooperation on a network level leading to low firing rates. (A)** Schematic of the connectivity of the simulated neural network. First, each neuron has one randomly chosen above-threshold post-synaptic and pre-synaptic connection (black), where supplemental connections are drawn with probability 0.1/N (purple), where N stands for the network size. Delays are selected randomly from $U(6, 9.5)$ ms and each neuronal $f_C$ is selected randomly from two values (light/dark green). **(B-E)** Normalized histograms of critical frequencies, $f_C$ (light blue), and firing frequencies (dark blue) obtained in simulations for the network topology **(A)** with N=2000. **(B)** $f_C$ is either 14.28 or 6.66 Hz. **(C)** $f_C$ is taken randomly from $U(6.66, 14.28)$ Hz. **(D)** $P_{fail}=0.07$ is added for all firing frequencies, even below $f_C$. **(E)** Spontaneous stimulations are added with an average rate of 1 Hz per neuron. All histograms were estimated several seconds after the initialization of the network (see MATERIALS AND METHODS), using a time window of about 50 s. The observed distribution, **(B-E)**, of the firing rates were found to be independent of the initial external stimulation patterns given to the network, indicating neuronal cooperation that reduces firing frequencies towards the lowest critical frequencies, $f_C$. All the results shown in this figure were produced in simulations.

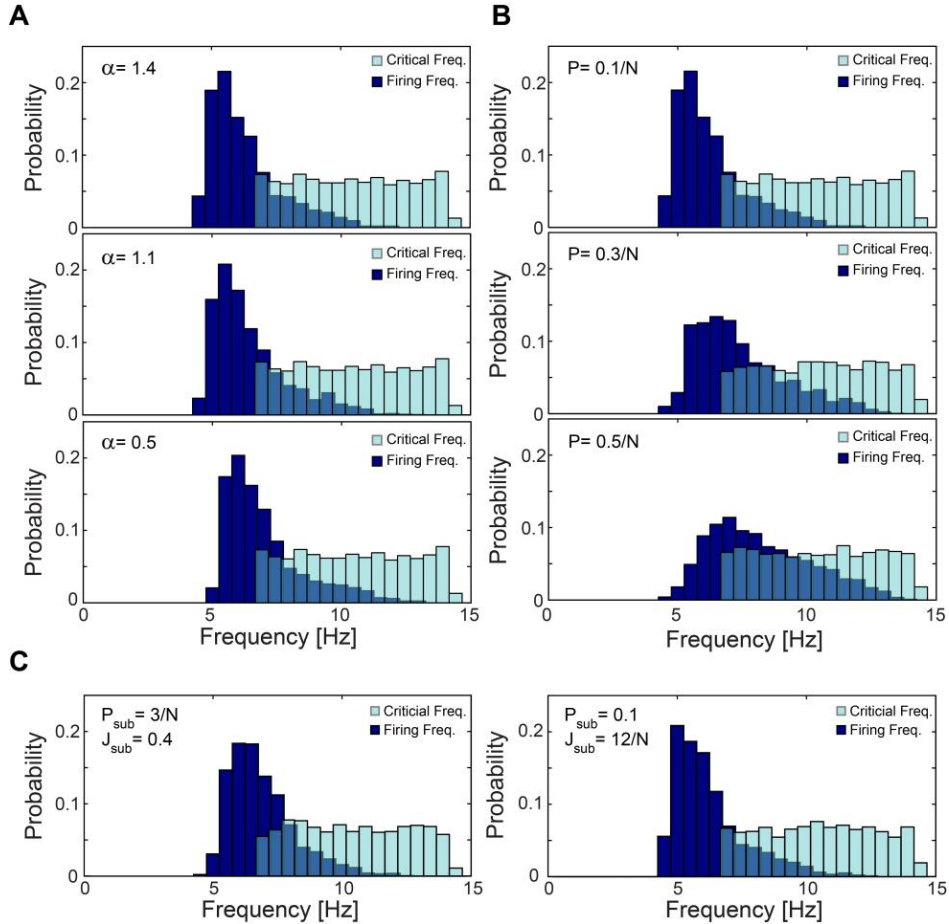

**FIGURE 12 | Robustness of low firing rates on a network level. (A-C)** The normalized histogram of critical frequency, $f_C$ (light blue), and firing frequency (dark blue). **(A)** The low firing activity is robust to variations in $\alpha$. An identical network topology as in **Figure 11C** but with different $\alpha$: 1.4 (upper panel), 1.1 (middle panel) and 0.5 (lower panel), with mean firing rate of 6.26, 6.43 and 6.96 Hz, respectively. **(B)** The low firing activity is robust to higher connectivity. Networks obeying the same statistical features as in **Figure 11C** but the probability for an additional above-threshold connection, P, (see MATERIALS AND METHODS) is 0.1/N (upper panel), 0.3/N (middle panel) and 0.5/N (lower panel), with mean firing rate of 6.26, 7.43 and 8.23 Hz, respectively. Results indicate that higher connectivity slightly increases the firing rates. **(C)** The low firing activity is robust to additional sparse/dense sub-threshold connectivity. Networks obeying the same statistical features as in **Figure 11C** where each neuron has, on the average, additional $N \cdot P_{sub}=3$ (left panel), $0.1 \cdot N$ (right panel) sub-threshold post-synaptic connections with the strength of $J_{sub}=0.4$ (left panel), 12/N (right panel). Note that the average total strength of the sub-threshold connections, per neuron, is preserved for both sparse and dense cases as $N \cdot P_{sub} \cdot J_{sub}=1.2$. All the results shown in this figure were produced in simulations.

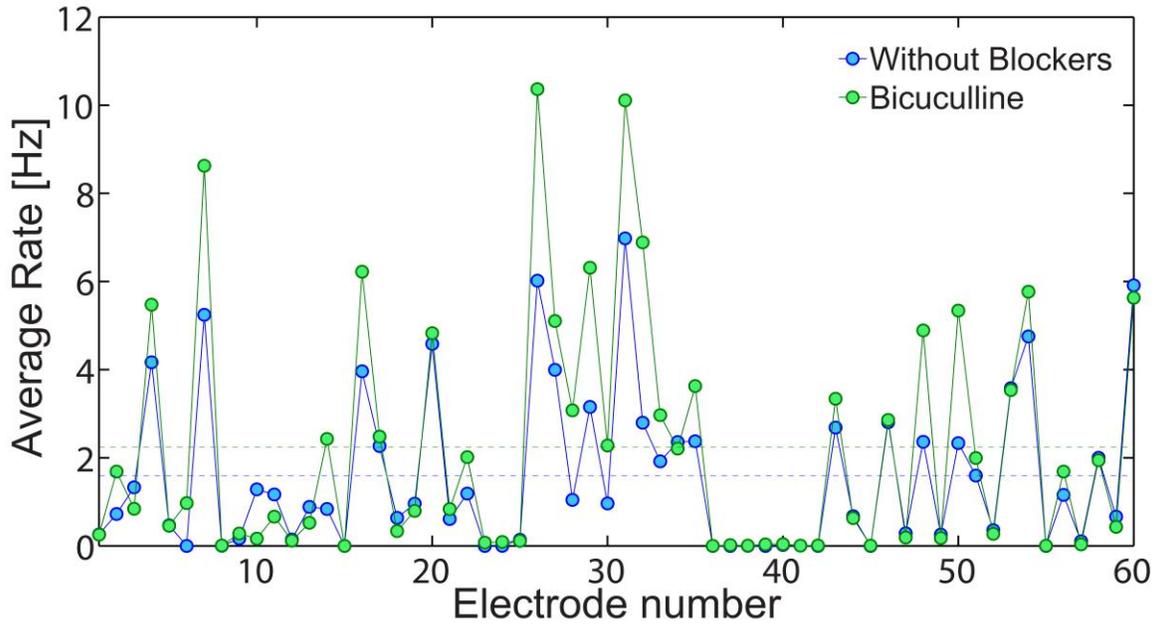

**Figure 13 | The effect of inhibition on the network spontaneous activity**. The average spontaneous spike detection rate recorded from all sixty electrodes of a plated MEA over 10 minutes where no external stimulations were given (blue). Similarly, the spontaneous activity of the same MEA with addition Bicuculline (5mM), which blocks inhibition (green). The average spike detection rates recorded from all electrodes (excluded the grounded 15[th] electrode) are presented by the dashed lines for each scenario.